# Wobbling of What?

Semenov D.A. (dasem@mail.ru)

International Research Center for Studies of Extreme States of the Organism at the Presidium of the Krasnoyarsk Research Center, Siberian Branch of the Russian Academy of Sciences

**Abstract.** A simple explanation for the symmetry of the genetic code has been suggested. An alternative to the wobble hypothesis has been proposed. The facts revealed in this study offer a new insight into physical mechanisms of the functioning of the genetic code.

The wobble hypothesis, which was first proposed more than 40 years ago [1], can explain two events: 1) formation of the uracil-guanine pair as a result of codon-anticodon interaction in the third position and 2) encoding of isoleucine by three codons (for the universal genetic code). This hypothesis is based on two statements: in the third position of the codon nucleotides can "wobble", and inosine can form pairs with uracil, cytosine, and adenine. Neither of these statements is necessary for explaining the abovementioned events, and the wobble hypothesis is just wrong. Once we admit this, we can get a better insight into the structure and functioning of the genetic code.

What makes the uracil-guanine pair possible? Uracil can exist either in the enol form or in the keto form. Uracil in the enol form can make two and even three hydrogen bonds with guanine, while uracil in the keto form would be able to make only one hydrogen bond. One can say that the enol form of uracil is stabilized by its interaction with the complementary guanine. The keto-enol tautomerism is a well-studied process and chemists will understand me without any further proofs.

The molecular basis of the wobble hypothesis has been criticized by other researchers [2], but I'd like to add a few comments.

Watson first considered nucleotide formulae in the enol form as the more probable, guided by reference books of that time [3]. Fortunately, I can give convincing arguments using commonly available information. Codon-anticodon interaction results in the formation of a short segment of the double helix. In the case of GC-rich codons, opposite to uracil in the third position of the codon there is inosine in the anticodone.

Crick's wobble hypothesis [1] allows a solution for this pair only by wobbling the third nucleotides of the codon and the anticodon. However, in the double helix, their position is also stabilized by stacking. If stacking exerts significant influence so that wobbling becomes impossible, then no uracil-inosine complementarity is possible for the keto form of uracil, i.e. there cannot be any hydrogen bonds. If in this case uracil is in enol form, in the state that most closely imitates cytosine, two hydrogen bonds can be established, without a nucleotide shift.

If C-I are Watson-Crick's pairs and U-I pairs are formed in accordance with the wobble hypothesis, it is not clear why these codons are always indistinguishable, although they have different conformations, which can be stabilized. Crick's hypothesis would be valid if cytosine and uracil were opposed by the same anticodon but as a component of different tRNAs, which would stabilize different conformations of the anticodon. This, however, is not so – the same tRNA always "confuses" cytosine with uracil in the third position of the codon. With the keto-enol tautomerism, both codons have the same conformation.

Why is uracil opposed by inosine in some anticodons and by guanine in others? According to Crick's hypothesis, inosine and guanine are indistinguishable to uracil. If we suppose the presence of the enol form of uracil, we can suggest that inosine emerged in anticodons in an evolutionary way because in this case the guanine amino group could not form a hydrogen bond. That is, in this case inosine (guanine without an amino group) is sufficient. Weakening and possible disruption of the hydrogen bond that has been formed involving the amino group must cause more than just binding of vacant hydrogen bonds to water molecules. Binding of two water molecules will transform the amino group from an active helper into an active obstacle.

I should add that the enol form of uracil can be registered in NMR spectra, which makes the necessary experiments easy to perform.

Incorporation of thymine into DNA may be accounted for as follows. Methylation could favor further stabilization of uracil in the keto form. Oxygen is an electron acceptor while the methyl group an electron donor, so the presence of the methyl group reduces the probability of the proton being near oxygen [4].

Let us now discuss the second part of Crick's hypothesis: why is isoleucine encoded by three codons? A more general question is: why are methionine and tryptophan encoded without degeneracy, by one codon each?

To answer this question, we have to refer to another study published in 1966 – Yu.B. Rumer's study of the symmetry of the genetic code table.

Let us consider the table of the genetic code letter doublets as it was proposed by Yu.B. Rumer [5, 6] (Table 1). The author focused his attention on the presence of the letter doublets (i.e. the first two nucleotides of the triplet, called "roots" by Yu.B. Rumer) and their ability or inability to encode just one amino acid. Of 16 letter doublets, 8 were strong (encoding just one amino acid) and 8 were weak (encoding more than one amino acid). Please note that this symmetry is characteristic of almost all dialects of the genetic code. The only exception has been comprehensively studied [7] and is accounted for by a mutation in aminoacyl-tRNA-synthetase, i.e. it is not related to the anticodon recognizing the codon.

|   | C | G | U | A |
|---|---|---|---|---|
| C | Pro | Arg | Leu | His Gln |
| G | Ala | Gly | Val | Asp Glu |
| U | Ser | Cys Trp/Stop | Phe Leu | Tyr Stop |
| A | Thr | Ser Arg | Ile Met | Asn Lys |

Table 1. The symmetry of the table of the letter doublets for the genetic code (according to Rumer). Strong letter doublets are marked in gray.

The keto-enol tautomerism is a good example of dramatic changes in the form and properties of certain nucleotides that occur under rather weak interactions.

A nucleotide triplet can be also assumed to change its form, but in this case a change in the form of the first two nucleotides occurs due to the third. Here I do not speak of the form of the triplet in the solution or the form of the triplet in mRNA, but rather the form of the double helix segment resulting from the codon-anticodon interaction.

This form is determined by two forces – complementary interactions and stacking (interaction of neighboring nucleotides). Stacking is a nonspecific event as there is a well-known formula: purine-purine>purine-pyrimidine>pyrimidine-pyrimidine.

In the third position of the codon the number of the formed hydrogen bonds is much less relevant than the nature of the bases – purine or pyrimidine. Thus, the presence of strong and weak letter doublets (Table 1) can naturally be related to the presence of stacking.

In codons with the CC, CG, GC, and GG letter doublets, the form of the doublet is only determined by complementary interaction. The three hydrogen bonds in each doublet make conformational changes impossible.

Let us now discuss the (UC-UG), (AC-AG), (CU-CA) and (GU-GA) letter doublets. In each doublet the first letter is strong and the second weak. The number of hydrogen bonds in each doublet is the same, but doublets with purine in the second position are the weak ones. Due to the presence of purine in the second position, two purines can occur one by one – in the second and the third positions of the codon. This construction imparts sufficient stress to the first two nucleotides to cause a change in their conformation. Please note that the anticodon is much less conformationally flexible because it is part of tRNA and is stabilized by its structure.

In codons with the UA and AA doublets, the conformation of the doublet can be changed due to the presence of purine in the third position of the codon. In the codons with the UU and AU doublets, the conformation of the codon can be changed by just a slight interaction between pyrimidine and purine.

The fact that complementary interactions are comparable with stacking in their effect may be surprising, but if this were not so and there were just one predominant type of nucleotide interactions, the other would not be described in textbooks. It is not less surprising that only half of the codon letter doublets are capable of conformational changes under the impact of stacking.

Rumer's symmetry [5, 6] can then be interpreted as follows: U>A because pyrimidine in the second letter of the codon prevents it from possible stacking-related conformational changes.

In the table of the universal genetic code there are two amino acids, each of which is encoded by one codon only: tryptophan (Trp), encoded by the UGG codon, and methionine (Met), encoded by AUG. This feature can be consistently explained based upon the same arguments that have been used to explain Rumer's symmetry. If there are letter doublets incapable of conformational changes when pyrimidine is replaced by purine in the last letter and there are letter doublets that readily change under these conditions, there must be letter doublets that are close to the equilibrium point. For these letter doublets even much smaller impacts may prove to be significant. This model can be illustrated "mechanically": one scale pan holds the total number of hydrogen bonds formed by the first two base pairs and the other – the stacking force between the second and the third nucleotides.

The UG doublet is situated on the table diagonal, i.e. it may be close to the equilibrium point. For its conformational change it may be important not only that the third position is occupied by purine but also that this is guanine. An additional hydrogen bond leads to a conformational change. The situation with the AU doublet is similar, but instead of the purine-purine interaction, we have to assume that its conformation can be changed by a weaker, pyrimidine-purine, interaction. In this case, the letter doublet itself is extremely prone to conformational changes: the doublet contains no nucleotides capable of forming three hydrogen bonds. The AU doublet is presumably close to the equilibrium point and the conformational transition is achieved at the expense of one hydrogen bond in the last nucleotide.

Thus, it is the first two nucleotides, or, to be more exact, the conformation of the first two nucleotides, that are the encoding part of the codon. For half of the letter doublets this conformation depends upon the last nucleotide, although it is this conformation rather than the whole sequence that is recognized. The number of different letter doublets of codons is not large: 8 (one for each strong doublet) + 16 (two for each weak one) = 24. There is only one codon for which all three nucleotides are significant – UGA, the stop codon. UGA's potential anticodon – UCA, placed into the tryptophan tRNA, would not be able to encode tryptophan. It would keep the conformation of the first two nucleotides unchanged. This can be a way to verify my speculations experimentally.

At the present time there are evidently no calculation procedures for molecule structure, which would take into account the influence of one hydrogen bond on the conformation of the double helix with three base pairs. This may be more significant than just a temporary technical limitation. For physical reasons, biological macromolecules can be in the state sensitive to weak changes, which accounts for their functional activity. Physics is familiar with such states; this is, e.g., the state near the critical point of the second-order phase transition. Theoretically, in these states, even very weak impacts can cause significant concerted structural changes. This is attractive for technology and even more attractive for evolution, which is originally sparing in expenditure of energy resources.

How could a similar effect be expressed in the functioning of the genetic code? Let us suppose that the codon that takes the most advantageous conformation makes the anticodon to accept the corresponding conformation. This effect cannot be strong as all our reasoning has been based on limited mobility of the anticodon. The effect must be certainly weaker than the effect of the formation of one hydrogen bond; however, if it caused a change in the conformation of the whole tRNA, this would suggest its active influence on protein biosynthesis. This would also suggest a crisis in contemporary computational methods in molecular biology: what is related to biological function of macromolecules would inevitably be lost due to calculation errors.

**Acknowledgement.**

The author would like to thank Krasova E. for her assistance in preparing this manuscript. I would like to thank Dr. Bystritsky A.A. for helpful discussions. I wish to express my gratitude to Dr. Wolfgang Pluhar for his appropriate suggestion and encourage.

**References:**
1. Crick F . Codon-anticodon pairing: the wobble hypothesis. *J Mol Biol* 19 (2): 548-55. (1966)
2. Pluhar W. The molecular basis of wobbling: An alternative hypothesys. J Theor. Biol. 169, 305-312. (1994)
3. Watson, J.D., *The Double Helix.* Atheneum, New York, 1968.
4. Zeegers-Huyskens Th. The bacity of the two carbonil bonds in uracil derivates. J. Mol. Struct. 198. 135-142. (1989)
5. Rumer Yu.B. On systematizing codons in the genetic code. *Doklady Akademii Nauk SSSR* 167, 6, 1394, (1966).
6. Rumer Yu.B. Systematizing of codons in the genetic code. *DAN SSSR.* 183, 1, 225-226, (1968).
7. Gomes, A., Costa, T., Carreto, L., Santos, M. On the molecular mechanism of the evolution of genetic code alterations. Molecular biology 40 (4): 564-569 (2006)